\documentclass[lettersize,journal]{IEEEtran}
\usepackage{array}
\usepackage[caption=false,font=footnotesize]{subfig}
\usepackage{textcomp}
\usepackage{stfloats}
\usepackage{url}
\usepackage{verbatim}
\usepackage{graphicx}
\usepackage{cite}
\usepackage[colorlinks=true, linkcolor=blue, citecolor=blue, urlcolor=blue]{hyperref}
\usepackage{booktabs}%提供命令\toprule、\midrule、\bottomrule
\usepackage{amsmath,amssymb,amsfonts}
\usepackage{graphicx}
\usepackage{textcomp}
\usepackage{xcolor}
\usepackage{subfig}
\usepackage{multirow}
\usepackage{upgreek}
\usepackage{color}
\usepackage{epstopdf}
\usepackage{bm}
\usepackage{array,color}
\usepackage[linesnumbered,ruled,vlined]{algorithm2e}
\usepackage[normalem]{ulem}
\usepackage{amsfonts}
\usepackage{graphics}
\usepackage{epsfig}
\usepackage{soul}
\usepackage{gensymb}
\soulregister\cite7
\soulregister\ref7
\usepackage{setspace}
\usepackage{svg}

\DeclareUnicodeCharacter{2212}{\ensuremath{-}}

\begin{document}

\title{Trajectory Design for UAV-Based Low-Altitude Wireless Networks in Unknown Environments: A Digital Twin-Assisted TD3 Approach}

\author{Jihao~Luo, Zesong~Fei,~\IEEEmembership{Senior Member,~IEEE}, Xinyi~Wang,~\IEEEmembership{Member,~IEEE}, Le~Zhao,
	
	 Yuanhao~Cui,~\IEEEmembership{Member,~IEEE},
	 Guangxu~Zhu,~\IEEEmembership{Member,~IEEE}, Dusit~Niyato,~\IEEEmembership{Fellow,~IEEE}
	%\vspace{-0.3cm}
	
	\thanks{Jihao Luo, Zesong Fei, Xinyi Wang and Le Zhao are with the School of Information and Electronics, Beijing Institute of Technology, Beijing 100081, China (e-mail: jihaoluo$\_$bit@icloud.com, feizesong@bit.edu.cn, wangxinyi@bit.edu.cn, tobin$\_$bit@icloud.com).}
	
	\thanks{Yuanhao Cui is with the School of Information and Communication Engineering, Beijing University of Posts and Telecommunications, Beijing 100876, China (e-mail: yuanhao.cui@bupt.edu.cn).}
	
	\thanks{Guangxu Zhu is with Shenzhen Research Institute of Big Data, The Chinese University of Hong Kong (Shenzhen), Shenzhen 518172, China (e-mail: gxzhu@sribd.cn).}
	
	\thanks{Dusit Niyato is with College of Computing and Data Science, Nanyang Technological University, Singapore (e-mail: dniyato@ntu.edu.sg).}
	\thanks{Corresponding author: Xinyi Wang.}
	
}

\markboth{Journal of \LaTeX\ Class Files,~Vol.~18, No.~9, August~2024}%
{Shell \MakeLowercase{\textit{et al.}}: A Sample Article Using IEEEtran.cls for IEEE Journals}


\maketitle

\begin{abstract}
Unmanned aerial vehicles (UAVs) are emerging as key enablers for low-altitude wireless network (LAWN), particularly when terrestrial networks are unavailable. In such scenarios, the environmental topology is typically unknown; hence, designing efficient and safe UAV trajectories is essential yet challenging. To address this, we propose a digital twin (DT)-assisted training and deployment framework. In this framework, the UAV transmits integrated sensing and communication signals to provide communication services to ground users, while simultaneously collecting echoes that are uploaded to the DT server to progressively construct virtual environments (VEs). These VEs accelerate model training and are continuously updated with real-time UAV sensing data during deployment, supporting decision-making and enhancing flight safety. Based on this framework, we further develop a trajectory design scheme that integrates simulated annealing for efficient user scheduling with the twin-delayed deep deterministic policy gradient algorithm for continuous trajectory design, aiming to minimize mission completion time while ensuring obstacle avoidance. Simulation results demonstrate that the proposed approach achieves faster convergence, higher flight safety, and shorter mission completion time compared with baseline methods, providing a robust and efficient solution for LAWN deployment in unknown environments.
\end{abstract}


\section{Introduction}

Unmanned aerial vehicles (UAVs) are revolutionizing the low-altitude economy by powering low-altitude wireless networks (LAWNs) that provide seamless and flexible connectivity for ground users (GUs) \cite{11045436, 10908560, 11152341, yuan2025ground}. With their three-dimensional (3-D) mobility, UAVs can rapidly extend connectivity in areas with limited infrastructure, during emergencies, or under dynamic network demands \cite{11053243, 11095407}. However, UAV-based LAWN deployment faces critical challenges, including limited onboard computation, collision risks, and fluctuating service requirements. To overcome them, efficient UAV trajectory design is crucial for ensuring reliable communication and safe operation in complex environments \cite{11150597, 10147371}.

Extensive efforts have been devoted to optimizing UAV trajectories, with a dominant focus on convex optimization frameworks. For example, Yuan \textit{et al.} \cite{9957134} proposed an iterative convex optimization framework to minimize mission execution time through efficient UAV trajectory design, demonstrating the potential of structured mathematical formulations for trajectory control. Complementary to this, Papaioannou \textit{et al.} \cite{10925885} utilized mixed-integer programming techniques to enhance trajectory planning, with a focus on maximizing the UAV’s coverage area. Beyond these model-based approaches, a variety of heuristic algorithms, such as genetic algorithms, particle swarm optimization, and ant colony optimization, have been explored for UAV path planning \cite{9286911, 10382648, 10546306}. However, despite their flexibility, both convex optimization and heuristic techniques face scalability bottlenecks and high computational overhead, which hinder their applicability in large-scale and dynamic network environments.

To overcome these constraints, deep reinforcement learning (DRL) has gained attention as it enables adaptive decision-making through trial-and-error interaction with the environment \cite{10239498, 10381761, 10925607}. DRL methods can be broadly categorized by action space. In discrete settings, value-based approaches such as deep Q-network (DQN) variants have been applied to UAV trajectory optimization. For instance, imitation-augmented DQN exploited channel maps from ground base stations to maximize sum rate \cite{9466945}, while double and dueling DQN frameworks improved energy efficiency and reduced information age in mobile edge computing (MEC) \cite{10706975} and vehicular networks \cite{10433247}. In continuous control tasks, policy-based and actor–critic methods provide finer trajectory adjustments. Examples include proximal policy optimization for enhancing spectrum cartography \cite{10097711}, soft actor-critic-based method for throughput fairness \cite{9815202}, and deep deterministic policy gradient (DDPG) for secrecy-rate maximization \cite{9801656}. Despite these advances, most DRL schemes rely on idealized simulations that fail to capture real-world uncertainties in LAWN deployment, especially when highly accurate reconstructions of physical conditions are not available. Training directly in physical environments remains costly and risky due to model instability, slow convergence, and potential collisions \cite{11126985}. Bridging this simulation-to-reality gap is thus a key challenge for practical DRL-driven UAV trajectory optimization.

The digital twin (DT) paradigm has recently been proposed to bridge this gap by constructing dynamic virtual environments (VEs) synchronized with the physical world through real-time and historical data integration \cite{mihai2022digital, 10989527, xu2023survey}. Open-source platforms such as Mago3D \cite{mago3d2023} and the Model Conductor-extended Framework \cite{shahsavari2021mcx} facilitate DT creation, enabling DRL agents to train within high-fidelity VEs and generate extensive experience, thereby accelerating convergence and improving robustness \cite{10078846}. UAV-specific studies have validated DT’s effectiveness in energy-efficient trajectory design \cite{zhao2023adaptive}, spectrum allocation \cite{luo2024gnn}, dynamic task assignment \cite{deng2023dynamic}, aerial MEC optimization \cite{tang2023digital}, and multi-UAV DRL acceleration \cite{tang2024digital}. However, applying DT to UAV deployment in unknown environments remains challenging, as no reliable prior data are available to initialize the DT. A critical open problem is how to construct a usable DT from scratch and progressively refine it with real-time sensing data collected during UAV operation \cite{11004029}. Such progressive updates are essential to maintain synchronization between the VE and the physical world, ensuring safe and efficient UAV deployment.

In this paper, we address the UAV trajectory design in unknown environments, where a UAV serves multiple GUs in a time-division multiple-access (TDMA) manner while simultaneously sensing and uploading environment information for DT construction. To accelerate model training, enhance deployment performance, and ensure flight safety, we propose a DT-assisted training and deployment framework (DTTDF), which integrates VE generation with real-time flight decision-making. Additionally, we propose a simulated annealing and twin-delayed deep deterministic policy gradient (TD3)-based trajectory design (SATD3TD) scheme, where simulated annealing provides efficient initial scheduling and TD3 ensures stable continuous control. By combining offline VE-driven training with online VE updates within the DTTDF, the DT-assisted SATD3TD model dynamically adapts UAV trajectories in real time, thereby improving both path efficiency and flight safety. The main contributions are outlined below.

The main contributions in this paper are summarized as follows.
\begin{enumerate}
	\item We propose a DTTDF for UAV deployment in completely unknown environments. During the training phase, the DT generates diverse realistic VEs to accelerate model convergence and improve learning robustness. In the deployment phase, the UAV continuously senses its surroundings to construct a real-time VE within the DT, which supports dynamic trajectory design, enhances decision accuracy, and ensures safe and efficient operation.
	\item We formulate the UAV trajectory design problem as a markov decision process (MDP), aiming to minimize mission execution time while satisfying flight safety constraints. We propose a SATD3TD scheme, which integrates user scheduling via simulated annealing with trajectory optimization via TD3, achieving efficient and safe UAV trajectory design.
	\item We integrate the SATD3TD algorithm into the DTTDF, proposing the DT-assisted SATD3TD scheme. Simulation results demonstrate that this approach accelerates model convergence, consistently ensures flight safety, and significantly reduces mission completion time compared with various baseline methods.
\end{enumerate}

The remainder of this paper is organized as follows. Section \ref{Sec.Systemmodel} introduces the mission environment, UAV model, and the communication model, as well as formulates the trajectory design problem. Section \ref{Sec.DTframework} describes the architecture and workflow of the proposed DTTDF and provides an overview of the DT-assisted SATD3TD process. Section \ref{Sec.Algorithm} presents the proposed algorithm in detail. Section \ref{Sec.Simulation} showcases the simulation results. Conclusions are drawn in Section \ref{Sec.Conclusion}.

\begin{figure}[t!]
	\centering
	\includegraphics[width=0.5\textwidth]{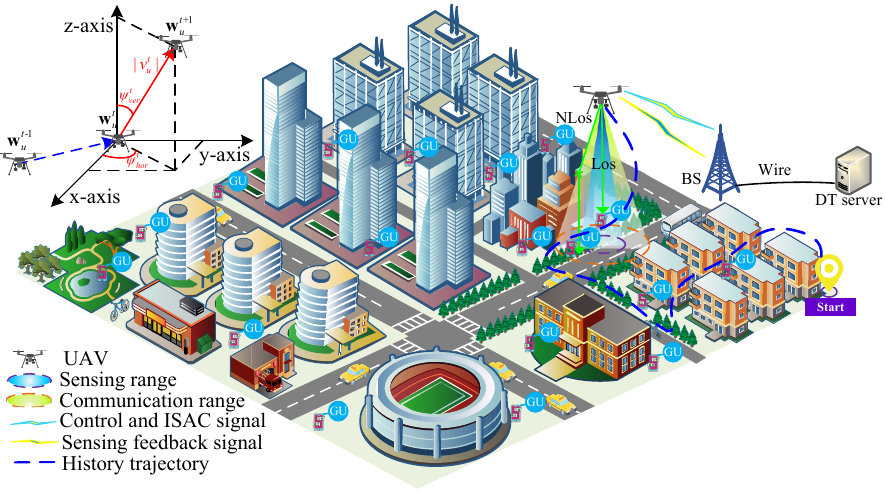}
	\caption{An illustration of UAV-based LAWN deployment system in an unknown environment. The UAV provides communication services to GUs while performing environment sensing, and the DT server leverages the sensing feedback to update the VE and generate flight policies. The top-left inset depicts the UAV’s movement and orientation in a 3D coordinate system.}
	\label{fig. system model}
	\vspace{0 cm}
\end{figure}

\section{System Model and Problem Formulation}\label{Sec.Systemmodel}
In this section, we present the system model for UAV-based LAWN deployment in unknown environments. We first describe the environment model, followed by the UAV motion model and the communication model, and finally formulate the trajectory design problem.

\subsection{Environment Model}\label{Environment Model}

As shown in Fig. \ref{fig. system model}, we consider a UAV-based LAWN deployment in an unexplored area. A ground BS first transmits communication data to a UAV, which relays it to GUs using an integrated sensing and communications (ISAC) waveform. This enables simultaneous downlink communication and environment sensing. The BS is connected via a wired link to a DT server, which continuously refines the VE based on UAV sensing data and updates UAV trajectory decisions in real time. The entire mission area is modeled as a 3-D square scene $\mathcal{E}_{ent}$, within which the building area $\mathcal{E}_{bui}$ represents the regions occupied by buildings. Additionally, a total of $K$ static GUs, denoted as $\mathcal{K} = \{1, 2, \ldots, K\}$, are randomly distributed on the ground across the area. The position of the $k$-th GU is expressed as $\mathbf{w}_k = [\bar{x}_k, \bar{y}_k, 0] \in \mathbb{R}^3_+$, where $\bar{x}_k$ and $\bar{y}_k$ are the fixed horizontal coordinates. The BS is assumed to know the GU positions in advance, allowing the DT server to perform more efficient scheduling and trajectory design. For convenience, Table \ref{abbreviations} summarizes the main assumptions used throughout this paper.

\subsection{UAV Motion Model}
The mission duration is divided into $T$ equal time slots, indexed by $t \in \mathcal{T} = \{1, 2, \ldots, T\}$, each of duration $\delta_s$ units. At time slot $t$, the UAV's position is represented by $\mathbf{w}^{t}_{u}=[x^t_u,y^t_u,z^t_u]\in \mathbb{R}^3_+$, where $x^t_u$ and $y^t_u$ indicate the time-varying horizontal coordinates, and $z^t_u$ signifies the time-varying vertical coordinate. The UAV updates its position from one time slot to the next based on a three-coordinates update rule, as illustrated in the upper-left part of Fig. \ref{fig. system model}. Specifically, the position at time slot $t+1$ is determined by its flight velocity, vertical flight angle, and horizontal flight angle at slot $t$, expressed as
\begin{equation}
	\begin{aligned}
		& x^{t+1}_u=x^{t}_u+v^{t}_u \cdot \sin(\psi^t_{ver}) \cdot \cos(\psi^t_{hor}) \cdot \delta_s, \\
		& y^{t+1}_u=y^{t}_u+v^{t}_u \cdot \sin(\psi^t_{ver}) \cdot \sin(\psi^t_{hor}) \cdot \delta_s, \\
		& z^{t+1}_u=z^{t}_u+v^{t}_u \cdot \cos(\psi^t_{ver})\cdot \delta_s,
	\end{aligned}
\end{equation}
where $v^{t}_u$ represents the average flight velocity at time slot $t$, $\psi^t_{ver}$ denotes vertical flight angle relative to the $z$-axis at time slot $t$, and $\psi^t_{hor}$ denotes the horizontal flight angle relative to the $x$-axis in the $xy$-plane at time slot $t$.

In order to enhance energy efficiency, the UAV employs a directional antenna with beamwidth $\beta$. This angle, combined with the UAV's altitude $z^t_u$ at time slot $t$, determines its coverage range as $d^t=z^t_u / \cos(\beta)$. Note that communication signals experience one-way propagation, while sensing echoes undergo round-trip propagation. Consequently, the communication coverage exceeds the sensing coverage, i.e., $\beta_{com}>\beta_{sen}$.

\begin{table}[t!]
	\caption{\normalsize Important assumption notations in this paper}
	\centering
	\renewcommand{\arraystretch}{1.0}
	\footnotesize
	\hspace{-0.7cm}
	\begin{tabular}{c|l}
		\toprule
		\multicolumn{1}{c|}{$\mathbf{Variable}$ $\mathbf{symbols}$} & \multicolumn{1}{c}{$\mathbf{Descriptions}$} \\
		\midrule
		%$L, W, H$ & The length, width, height of the environment \\
		$\mathcal{E}_{ent}, \mathcal{E}_{bui}$ & The sets of mission areas and building areas \\
		$\boldsymbol{\psi}_{ver}, \boldsymbol{\psi}_{hor}$ & The sets of the UAV's vertical and horizontal angles \\
		$\mathcal{K}$ & The set of the GUs \\
		$\mathcal{T}$ & The set of the mission slots \\
		$\boldsymbol{v}_u$ & The set of velocities of the UAV \\
		%$\mathbf{w}^{0}_{u}$ & The start position of the UAV \\
		$\mathbf{w}^{t}_{u}$ & The position of the UAV at slot $t$ \\
		$\mathbf{w}_k$ & The position of GU $k \in \mathcal{K}$ \\
		$\mathbf{g}_s$ & The user scheduling vector \\
		$d^t_{sen}$ & The sense range for the UAV at slot $t$ \\
		$d^t_{com}$ & The communication range for the UAV at slot $t$ \\
		$\Lambda_k$ & The throughput requirement for GU $k \in \mathcal{K}$ \\
		$t_k$ & The min slots required for the UAV to serve GU $k$ \\
		$h^t_{b,u}$ & The channel gain of BS-UAV link at slot $t$\\
		$h^t_{u,k}$ & The channel gain of UAV-GU link at slot $t$\\
		$R^t_{b,u}$ & The transmission rate of BS-UAV link at slot $t$ \\
		$R^t_{u,k}$ & The transmission rate of UAV-GU link at slot $t$ \\
		$R^t_{\text{eff},k}$ & The effective transmission rate to GU $k$ at slot $t$ \\
		$M^t_{u,k}$ & The transmission data volume to GU $k$ at slot $t$ \\
		$l^t_{u,k}$ & The type of UAV-GU link at slot $t$\\
		\bottomrule
	\end{tabular}
	\vspace{0 cm}
	\label{abbreviations}
\end{table}

\subsection{Communication Model}

\begin{figure}[htbp]
	\centering
	%\hspace{0.9cm}
	\includegraphics[width=0.50\textwidth]{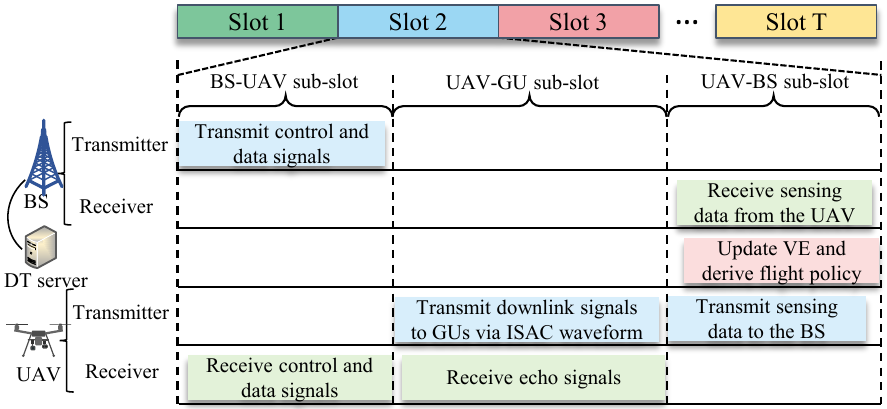}
	\caption{An illustration of the time slot protocol for the mission. Each slot is divided into three sub-slots that complete a full cycle of operations, namely BS-UAV control and data transmission, UAV-GU service with ISAC signals, and UAV-BS sensing feedback, thereby enabling both communication and DT-driven trajectory updates.}
	\label{fig. subslot}
	\vspace{0 cm}
\end{figure}

The UAV serves the GUs in a TDMA manner. To accommodate the UAV relay communication tasks, each slot is further divided into three sub-slots, including BS-UAV sub-slot, UAV-GU sub-slot, and UAV-BS sub-slot, as illustrated in Fig. \ref{fig. subslot}. The durations of these sub-slots are $\delta_1$, $\delta_2$, and $\delta_3$, respectively, satisfying $\delta_1 + \delta_2 + \delta_3 = \delta_s$. Furthermore, we assume that the first two sub-slots have equal duration, i.e., $\delta_1=\delta_2$. The functions of the three sub-slots are summarized below.
\begin{enumerate}
	\item BS-UAV sub-slot: The BS transmits control instructions and data signals to the UAV, and the UAV receives these signals.
	\item UAV-GU sub-slot: The UAV forwards the received data signals to the GUs using an ISAC waveform, while simultaneously collecting echoes for environmental sensing. 
	\item UAV-BS sub-slot: The UAV uploads the sensed data to the DT server through the BS, enabling VE updates and the derivation of a new flight policy for the next slot.
\end{enumerate}

Since the UAV operates as a relay, the effective communication rate is determined by the lower rate between the BS–UAV and UAV–GU links. To reflect realistic propagation, both large-scale path loss and small-scale fading are incorporated into the channel models. For the BS-UAV link, a line-of-sight (LoS) condition is assumed owing to the elevated position of the BS and the UAV’s altitude. The large-scale fading for the BS-UAV link at time slot $t$, denoted as $\mathrm{LSF}^t_{b,u}$, is modeled as
\begin{equation}
	\mathrm{LSF}^t_{b,u}[\mathrm{dB}] = \mathrm{PL}^t_{b,u} + \gamma_{\mathrm{LoS}},
\end{equation}
where $\gamma_{\mathrm{LoS}}$ represents shadowing in the LoS link, and the path loss, $\mathrm{PL}^t_{b,u}$, is given by
\begin{equation}
	\mathrm{PL}^t_{b,u} [\mathrm{dB}] = 20\log_{10}(d^t_{b,u}) + 20\log_{10}(f_c) + 20\log_{10}\left(\frac{4\pi}{c}\right),
\end{equation}
where $d^t_{b,u}= \Vert \mathbf{w}_{b} - \mathbf{w}^{t}_{u} \Vert$ denotes the Euclidean distance between the BS positioned at $\mathbf{w}_{b}$ and the UAV at slot $t$, $f_c$ represents the carrier frequency, and $c$ denotes the speed of light. The small-scale fading, $\mathrm{SSF}^t_{b,u}$,  captures rapid channel fluctuations. Thus, the channel gain, $h^t_{b,u}$, is expressed as
\begin{equation}
	h^t_{b,u} = 10^{-\frac{\mathrm{LSF}^t_{b,u}}{20}} \mathrm{SSF}^t_{b,u}.
\end{equation}
The transmission rate of the BS-UAV link is then given by
\begin{equation}
	R^t_{b,u} = B \log_2 \left(1 + \frac{p_b |h^t_{b,u}|^2}{N_0}\right),
\end{equation}
where $B$ denotes the available spectrum bandwidth, $p_b$ represents the constant BS transmit power, and $N_0$ stands for the power of zero-mean additive white Gaussian noise (AWGN).

For the UAV-GU link, the air-to-ground (A2G) channel is categorized as either LoS or non-line-of-sight (NLoS) based on the real-world topology constructed by the DT server, the reported GU positions, and the UAV's position. The communication range is defined as $d^t_{com}=z^t_u / \cos(\beta_{com})$. If the distance between the UAV and GU $k$ exceeds $d^t_{com}$, no links exists. Otherwise, the DT server evaluates whether the link $l^t_{u,k}$ is obstructed by buildings, assigning $l^t_{u,k}=1$ for LoS and $l^t_{u,k}=0$ for NLoS. The large-scale fading for the UAV-GU channel is expressed as
\begin{equation}
	\mathrm{LSF}^t_{u,k}[\mathrm{dB}]=
	\begin{cases} 
		\mathrm{PL}^t_{u,k} + \gamma_{\mathrm{LoS}}, & \text{if } l^t_{u,k} = 1, \\
		\mathrm{PL}^t_{u,k} + \gamma_{\mathrm{NLoS}}, & \text{if } l^t_{u,k} = 0, \\
	\end{cases}
\end{equation}
where $\gamma_{\mathrm{NLoS}}$ accounts for shadowing in NLoS links, and the path loss is calculated by
\begin{equation}
	\mathrm{PL}^t_{u,k} [\mathrm{dB}] = 20\log_{10}(d^t_{u,k}) + 20\log_{10}(f_c) + 20\log_{10}\left(\frac{4\pi}{c}\right).
\end{equation}
Here, $d^t_{u,k}= \Vert \mathbf{w}^{t}_{u} - \mathbf{w}_k \Vert$ denotes the Euclidean distance between the UAV and the $k$-th GU at time slot $t$. The small-scale fading, $\mathrm{SSF}^{t}_{u,k}$, accounts for rapid channel variations. The channel gain is calculated by
\begin{equation}
	h^t_{u,k} = 10^{-\frac{\mathrm{LSF}^t_{u,k}}{20}} \mathrm{SSF}^t_{u,k}.
\end{equation}
The UAV-GU transmission rate is given by
\begin{equation}
	R^t_{u,k} = B \log_2 \left(1 + \frac{p_u |h^t_{u,k}|^2}{N_0}\right),
\end{equation}
where $p_u$ denotes the UAV's transmit power \footnote{In the single-UAV scenario without inter-UAV interference and TDMA-based GU scheduling, it is reasonable to set $p_u$ to a constant maximum value.}.

Therefore, the effective transmission rate to GU $k$ at time slot $t$ is calculated by
\begin{equation}
	R^t_{\text{eff},k} = \min \left\{R^t_{b,u}, R^t_{u,k}\right\}.
\end{equation}
The data volume transmitted to GU $k$ in time slot $t$ is
\begin{equation}
	M^t_{u,k} = R^t_{\text{eff},k} \cdot \delta_2.
\end{equation}
To satisfy the throughput requirement $\Lambda_k$ for GU $k$, the minimum number of time slots, denoted as $t_k$, that the UAV needs to serve GU $k$ must satisfy
\begin{equation}
	\sum_{t=1}^{t_k} M^t_{u,k} \geq \Lambda_k.
\end{equation}
The mission is considered completed once the UAV has transmitted the required data volume to all GUs. Better channel quality increases $R^t_{\text{eff},k}$, allowing more data per slot and reducing the total slots needed, thereby shortening mission time.

\subsection{Problem Formulation}\label{Problem Formulation}
In this paper, we aim to optimize the UAV's trajectory to minimize the mission completion time while ensuring obstacle avoidance, which can be formulated as
\begin{equation}
	\begin{alignedat}{2}
		\min_{\boldsymbol{v}_u, \boldsymbol{\psi}_{ver}, \boldsymbol{\psi}_{hor}} \quad & \sum_{k=1}^K t_k \cdot \delta_s, \\
		\text{s.t.} \quad & C1: \ 0 \leq v^t_u \leq v_{max},\ \forall t \in \mathcal{T}, \\
		& C2: \ 0 \leq \psi^t_{ver} \leq \pi,\ \forall t \in \mathcal{T}, \\
		& C3: \ 0 \leq \psi^t_{hor} \leq 2\pi,\ \forall t \in \mathcal{T}, \\
		& C4: \ \mathbf{w}^{t}_{u} \in \mathcal{E}_{ent},\ \forall t \in \mathcal{T}, \\
		& C5: \ \mathbf{w}^{t}_{u} \notin \mathcal{E}_{bui},\ \forall t \in \mathcal{T}, \\
		& C6: \ \sum_{t=1}^{t_k} M^t_{u,k} \geq \Lambda_k,\ \forall k \in \mathcal{K},
	\end{alignedat}
\end{equation}
where $\boldsymbol{v}_u=\{v^1_u,v^t_u,\ldots,v^t_u\}$ represents the set of UAV's velocities, while $\boldsymbol{\psi}_{ver}=\{\psi^1_{ver},\psi^2_{ver},\ldots,\psi^T_{ver}\}$ and $\boldsymbol{\psi}_{hor}=\{\psi^1_{hor},\psi^2_{hor},\ldots,\psi^T_{hor}\}$ denote the sets of vertical and horizontal flight angles, respectively. Constraints $C1$-$C3$ limit the velocity, vertical angle and horizontal angle of the UAV, respectively, with $v_{max}$ denoting the maximum velocity. $C4$ and $C5$ ensure that the UAV remains within the designated mission area and avoids colliding with buildings. $C6$ guarantees that the cumulative data transmitted to each GU meets the minimum required data volume. 

\begin{figure*}[t]
	\centering
	\includegraphics[width=\textwidth]{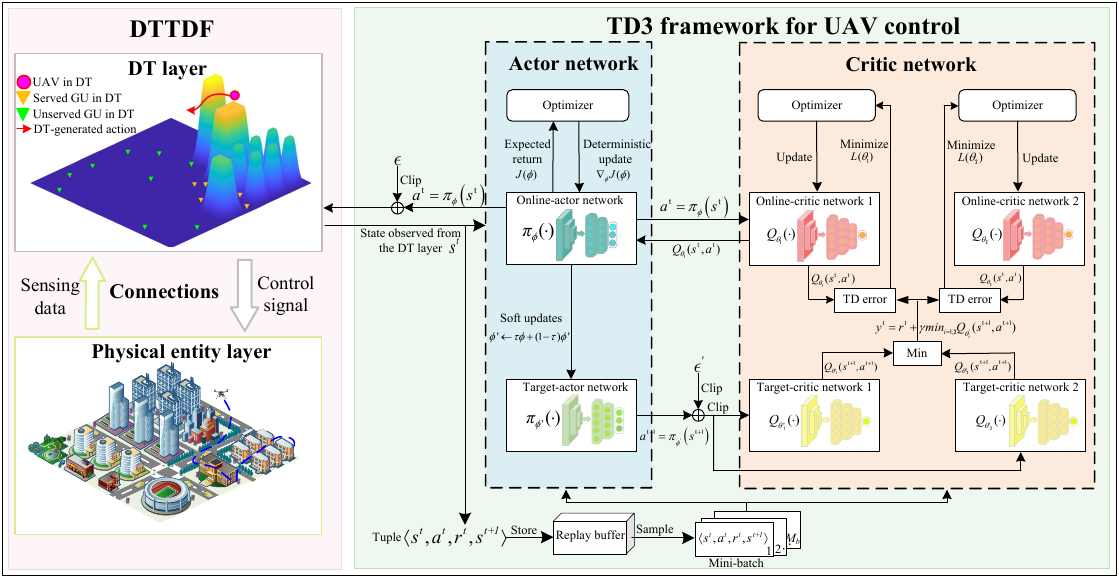}
	\caption{An illustration of the DTTDF and UAV control mechanism based on the TD3 framework. The left side shows the DTTDF, where UAV sensing data update a VE in the DT layer, which then generates control signals to guide the UAV’s trajectory. The right side depicts the TD3 framework, which leverages state observations from the DT to derive the UAV’s flight policy via twin actor-critic networks, experience replay, soft target updates, etc.}
	\label{fig:TD_TD3}
	\vspace{0cm}
\end{figure*}

\section{DTTDF Triggered Dynamic Trajectory Design}\label{Sec.DTframework}
In this section, we introduce the proposed DTTDF, detailing its core components, DT-assisted model training and mission execution process, and the integrated user scheduling and dynamic trajectory design strategy.

\subsection{Components of DTTDF}
The proposed DTTDF is illustrated in Fig. \ref{fig:TD_TD3}, consisting of three primary components: the physical entity layer, the DT layer, and their connections. The physical entity layer represents the real-world mission scenario, the DT layer functions as its virtual counterpart, and the connections serve as the bidirectional interaction between them. Each component is explored in detail as follows.
\begin{enumerate}
	\item Physical entity layer: This layer comprises the physical UAV, the mission execution area, and the GUs awaiting service. Owing to its limited computational and storage capacity, the UAV cannot train the SATD3TD model or store the vast sensing data generated during the mission. Instead, it functions purely as an agent, observing the environment, executing control actions received from the BS, delivering communication services to GUs, and uploading sensing data to the DT server.
	\item DT layer: Running on the DT server with abundant computational and storage resources, this layer constructs and maintains a high-fidelity virtual replica of the physical entities. The VE captures the UAV’s mobility, GU distribution, and environmental features such as buildings. Based on this VE, the SATD3TD model is trained and executed in the DT layer, enabling the generation of effective and safe control actions for guiding the UAV.
	\item Connections: Bidirectional links between the physical and virtual layers ensure synchronization and coordination. The UAV continuously uploads sensing data, which allows the DT layer to incrementally refine the VE. In return, the DT layer generates actions through the trained SATD3TD model and transmits them to the UAV via the BS for execution. To guarantee reliable and timely consistency between the UAV and its virtual counterpart, sufficient communication resources must be allocated to maintain stable links \cite{cakir2023synchronize}.	
\end{enumerate}

\subsection{DT-assisted Model Training and Mission Execution}
Since the mission area's topology is unknown in advance, directly training a model with real scenario data is challenging, and unknown buildings may threaten the UAV's safety. To address this, we adopt the DT technique, which enables offline model training before mission execution and provides continuous safety assurance during flight.

In the training phase, the DT server generates multiple VEs, allowing the virtual UAV to interact with diverse simulated mission scenarios. These interactions produce MDP-based experiences, which are stored in a replay buffer to train the SATD3TD model. With the DT server’s abundant computational resources, the training process is significantly accelerated, leading to a well-optimized model that is subsequently deployed to the virtual UAV to enhance its decision-making capability.

During mission execution, the DT server continuously monitors UAV actions and evaluates potential safety risks using the constructed VE. Before each action is transmitted to the UAV, it is checked for predicted collisions with obstacles or boundary violations. Unsafe actions are immediately canceled, and the UAV is instructed to remain stationary for that slot. This closed-loop verification ensures that UAV operations remain safe while following the guidance of the trained SATD3TD model. 

\subsection{User Scheduling and Trajectory Design in SATD3TD}
The DT server performs the trajectory design for the UAV deployment within the DTTDF in two stages: initial user scheduling and dynamic trajectory design. In the first stage, since the positions of GUs are known in advance whereas the positions of buildings remain initially unknown, the DT server employs a simulated annealing-based algorithm to derive an initial user schedule without considering building locations.

Subsequently, in the dynamic trajectory design stage, the UAV follows the initial schedule while continuously adapting its trajectory through the TD3 model based on real-time observations. As the UAV explores the environment, it detects previously unknown buildings and temporarily adjusts its path to avoid collisions through TD3. If such adjustments disrupt the original service order, the UAV prioritizes the nearest unserved GU within its communication range. Once the obstacle is bypassed, the UAV resumes the initial schedule, giving precedence to unserved GUs earlier in the list. By leveraging TD3 for dynamic decision-making, the UAV achieves a balance between safe navigation and service efficiency, thereby minimizing mission completion time in complex, previously unknown environments.

\section{The Proposed Trajectory Design Mechanism}\label{Sec.Algorithm}

\subsection{Simulated Annealing-based User Scheduling}\label{user scheduling}
To leverage the known positions of GUs and derive an efficient user schedule, we propose a simple yet effective user scheduling mechanism based on simulated annealing, as detailed in Algorithm \ref{alg:SA_algorithm}. The algorithm takes the UAV’s initial position, and the positions of all GUs as input, producing an optimized scheduling vector, $\mathbf{g}_{s}$, as output. Unlike conventional approaches, which is prone to local optima and requires extensive iterations to converge \cite{rajwar2023exhaustive}, our approach leverages problem-specific adaptations for UAV deployment. The detailed procedure of Algorithm \ref{alg:SA_algorithm} is elaborated below.

\subsubsection{\textbf{Initialization based on Nearest Greedy}}
Instead of adopting a random initial solution as in classical scheme, we initialize the process using a nearest greedy strategy. Starting from the UAV's initial position, the algorithm iteratively selects the closest unserved GU from the remaining GU set $\mathcal{G}r$ to construct a greedy scheduling vector, $\mathbf{g}_{g}$ (lines 1–5). This initialization provides a high-quality starting point and accelerates convergence.

\subsubsection{\textbf{Scheduling Refinement}}
Building on the greedy initialization, the refinement phase uses $\mathbf{g}_{g}$ as the initial schedule, with its total service distance set as both the initial and minimum values (line 6). At each iteration, a candidate solution is generated by applying a neighborhood operator to perturb the current schedule (lines 8–9). We employ multiple neighborhood operators, $\mathcal{O}=\{\text{swap}, \text{insert-move}, \text{reverse-subsequence} \}$, with one operator $o\in \mathcal{O}$ selected based on a temperature-dependent probability to balance exploration and exploitation: larger perturbations (e.g., reverse) are favored at high temperatures, while smaller ones (e.g., swap) prevail at low temperatures. The probability is defined as
$P(o \mid T) = \frac{\beta_o \exp(\alpha_o T)}{\sum_{o' \in \mathcal{O}} \beta_{o'} \exp(\alpha_{o'} T)},$ where $\beta_o$ and $\alpha_o$ denote the bias and sensitivity coefficient of operator $o$, respectively. The objective function is defined as the total UAV travel distance to serve all GUs, which directly reflects service efficiency. The candidate is accepted if it improves the distance, or with a probability determined by the Metropolis criterion, which decreases as the temperature gradually cools (lines 10–12). Whenever a shorter distance is found, the best-known schedule is updated (lines 13–14). By integrating greedy initialization with adaptive neighborhood exploration, the proposed algorithm offers a lightweight yet efficient solution for user scheduling in the DTTDF.

\begin{algorithm}[t!]
	\setstretch{0.9}
	\SetKwData{Left}{left}\SetKwData{This}{this}\SetKwData{Up}{up}
	\SetKwFunction{Union}{Union}\SetKwFunction{FindCompress}{FindCompress}
	\SetKwInOut{Input}{Input}\SetKwInOut{Output}{Output}\SetKwInOut{Initialize}{Initialize}
	
	\Input{UAV start position $\mathbf{w}^{0}_u$, and positions of all GUs $\mathbf{w}_k$, $\forall k \in \mathcal{K}$}
	\Output{User scheduling vector, $\mathbf{g}_{s}$}
	\Initialize{Temperature $T=T_{0}$, cooling rate $C_{r}$, and maximum iteration number $I^{SA}_{max}$}
	
	Set greedy user scheduling vector $\mathbf{g}_{g} \leftarrow \emptyset$, the UAV current position $\mathbf{w}^c_{u} \leftarrow \mathbf{w}^{0}_u$, total distance $d_t\leftarrow 0$, and remaining GU set $\mathcal{G}_r \leftarrow \{(k, \mathbf{w}_k) \mid k \in \mathcal{K}\}$\;
	\While{$\mathcal{G}_r$ is not empty}{
		Find $(k^*, \mathbf{w}_{k^*}) \in \mathcal{G}_r$ such that distance $d(\mathbf{w}^c_{u}, \mathbf{w}_{k^*})$ is minimized\;
		$d_t = d_t+d(\mathbf{w}^c_{u}, \mathbf{w}_{k^*})$\;
		Add $k^*$ to $\mathbf{g}_{g}$, Remove $(k^*, \mathbf{w}_{k^*})$ from $\mathcal{G}_r$, $\mathbf{w}^c_{u} \leftarrow \mathbf{w}_{k^*}$\;
	}
	Set user scheduling vector $\mathbf{g}_{s} \leftarrow \mathbf{g}_{g}$, simulated annealing distance $d_{s} \leftarrow d_t$, and minimum distance $d_{min} \leftarrow d_{t}$\;	
	\For{$i = 1$ \KwTo $I^{SA}_{max}$}{
		$\mathbf{g}_{new} \leftarrow \text{SelectOperatorByTemp}(\mathbf{g}_{g}, T)$\;
		$d_{new} \leftarrow  \text{CalcuDist}(\mathbf{w}^{0}_u, \mathbf{g}_{new})$\;
		$r \leftarrow \text{Random}(0,1)$\;
		\If{$d_{new} < d_{s}$ \text{ or } $r < \exp((d_{s} - d_{new}) / T)$}{
			$\mathbf{g}_{g} \leftarrow \mathbf{g}_{new}$, $d_{s} \leftarrow d_{new}$\;
		}
		\If{$d_{s} < d_{min}$}{
			$\mathbf{g}_{s} \leftarrow \mathbf{g}_{g}$, $d_{min} \leftarrow d_{s}$\;
		}
		$T \leftarrow T \times C_{r}$\;
	}
	\Return $\mathbf{g}_{s}$\;
	\caption{Simulated annealing-based user scheduling }
	\label{alg:SA_algorithm}
\end{algorithm}

\subsection{MDP Formulation for UAV Deployment}
Based on the system model established in Section \ref{Sec.Systemmodel}, we formulate the optimization problem as an MDP, with the UAV acting as the agent, aiming to maximize the cumulative reward associated with mission execution time and collision avoidance. Typically, an MDP is represented by the tuple $\langle \mathcal{S}, \mathcal{A}, \mathcal{R}, P \rangle$, which characterizes the agent’s sequential decision-making process. Here, $\mathcal{S}$, $\mathcal{A}$, $\mathcal{R}$, and $P$ denote the state space, action space, reward function, and state transition probability, respectively. The following elaboration provides detailed insights of these four components. 

\subsubsection{\textbf{State space}}
\begin{figure}[htbp]
	\centering
	\hspace{-0.5cm}
	\subfloat[$\mathbf{S}^t_1$]{
		\includegraphics[width=0.24\textwidth]{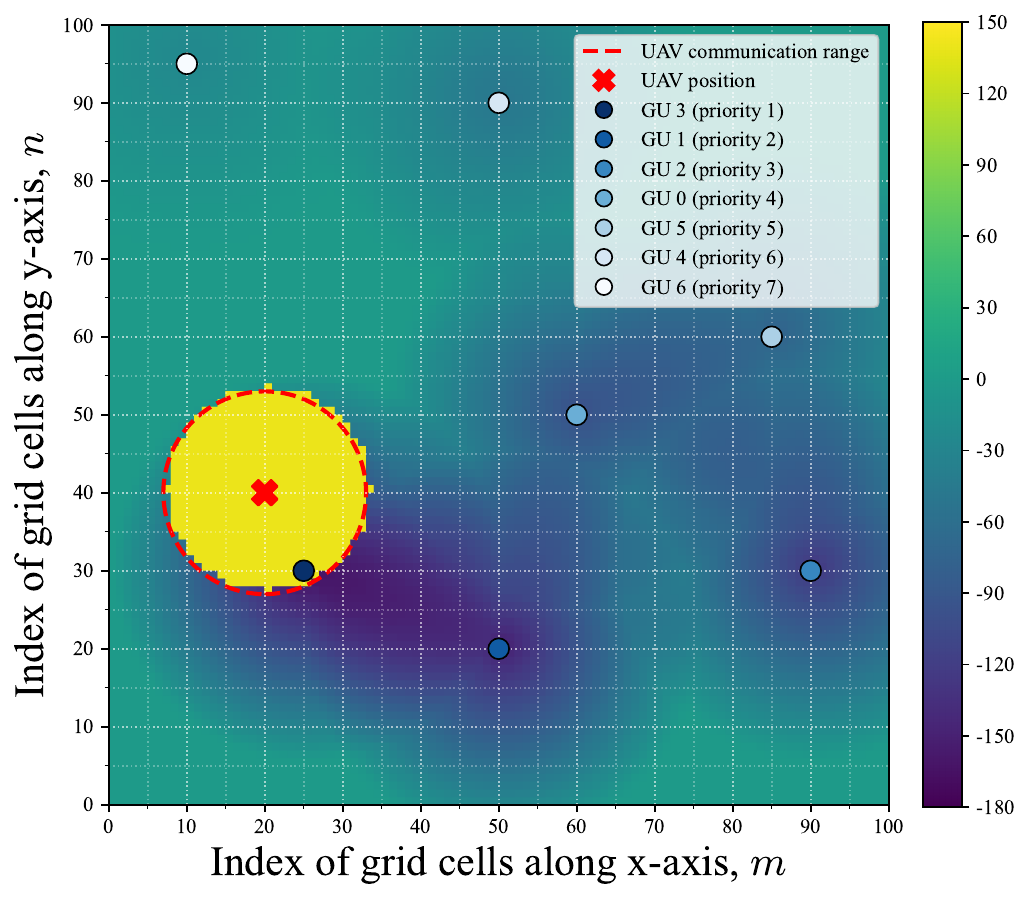}}
	\hspace{0\textwidth} 
	\subfloat[$\mathbf{S}^t_2$]{
		\includegraphics[width=0.24\textwidth]{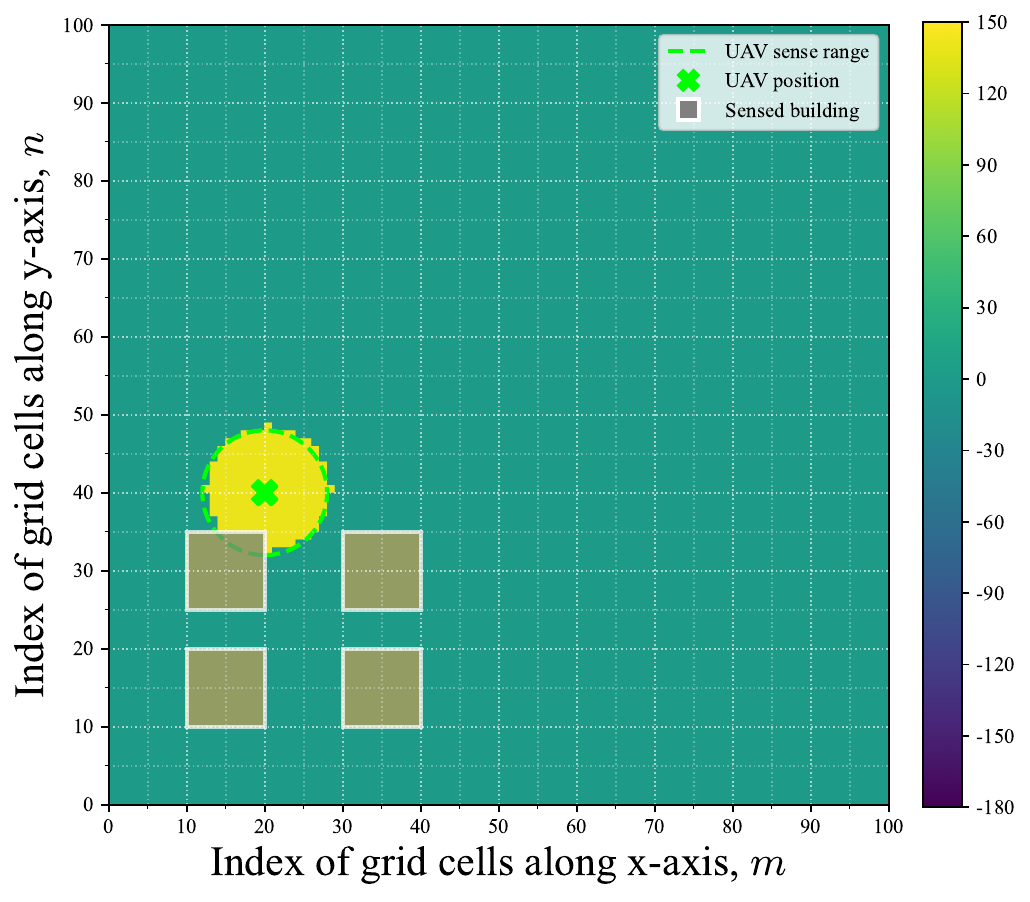}}
	\caption{An illustration of the state space with two components. 
		(a) $\mathbf{S}^t_1$ for service delivery, where the APF attracts the UAV toward GUs according to their priority. The UAV’s communication range is indicated by the dashed circle. 
		(b) $\mathbf{S}^t_2$ for obstacle avoidance, showing detected buildings along with the UAV’s position and sensing range, providing information for safe navigation.}
	\label{fig. state}
\end{figure}

The state space $\mathcal{S}$ includes all possible states $s^t$ at each time slot $t$. To enable the agent to learn a robust policy, we designed a tailored state representation that integrates essential mission information. Each state $s^t$ is defined by the positions of the UAV and GUs, detected buildings, and the user scheduling policy derived from Section \ref{user scheduling}. To simultaneously address the objectives of providing communication services to all GUs and avoiding obstacle collisions, we define the state at each time slot as a combination of two components, denoted as $s^t \triangleq [\mathbf{S}^t_1,\mathbf{S}^t_2]$. Each component is represented by an $m \times n$ matrix, where $m$ and $n$ correspond to the index of grid cells used to discretize the mission area. These components are specified as
\begin{enumerate}
	\item $\mathbf{S}^t_1$: This matrix facilitates the UAV’s delivery of communication services to GUs based on the user scheduling vector. Inspired by \cite{battocletti2021rl}, we employ artificial potential field (APF) techniques to integrate user scheduling with GU position information. The APF method generates a virtual force field, a mathematical construct that simulates attractive forces to guide the UAV’s motion. In this field, each GU generates an attractive potential, pulling the UAV along the negative gradient toward its position. We design the attractive field to radiate from each GU’s position, with its potential increasing linearly with distance within a specified range. Specifically, the potential at an arbitrary position $\mathbf{w}$ is defined as	
	\begin{equation}
		U_{att}(\mathbf{w})=
		\begin{cases} 
			\mu_k \cdot (d_{w_k} - J_{d,p}), & \text{if } d_{w_k} \leq J_{d,p} , \\
			0, & \text{if } d_{w_k} > J_{d,p}, \\
		\end{cases}
	\end{equation}
	where $\mu_k $ is a positive constant representing the attractive strength of GU $k$,  $d_{w_k}=\Vert \mathbf{w}_k - \mathbf{w} \Vert$ denotes the distance between $\mathbf{w}$ and $\mathbf{w}_k$, and $J_{d,p}$ represents the distance threshold for the potential. To prioritize GUs earlier in the schedule, we sort $\mu_k$ in descending order according to the user scheduling vector, ensuring the UAV serves higher-priority GUs first. The UAV is represented in the matrix as a circle centered at its position, with a radius of $d^t_{com}$ and a value of $z^t_u$. An example of $\mathbf{S}^t_1$ is illustrated in Fig. \ref{fig. state}(a).	
	\item $\mathbf{S}^t_2$: This matrix supports the UAV in avoiding obstacles by integrating its position and information about detected buildings. Specifically, the UAV is represented as a circle centered at its position, with a radius of $d^t_{sen}=z^t_u / \cos(\beta_{sen})$ and a value of $z^t_u$. Detected buildings are discretized in the matrix according to their sensed locations, with values corresponding to their sensed heights. An example of $\mathbf{S}^t_2$ is shown in Fig.~\ref{fig. state}(b).
\end{enumerate}

\subsubsection{\textbf{Action space}}
The action space $\mathcal{A}$ comprises all feasible actions $a^t$ at each time slot $t$. Each action represents the UAV’s flight decision, comprising its velocity, vertical flight angle, and horizontal flight angle, expressed as
\begin{equation}
	a^t=\{v^t_u, \psi^t_{ver}, \psi^t_{hor}\}.
\end{equation}

\subsubsection{\textbf{Reward function}}
The reward function $\mathcal{R}$ generates a reward $r^t$ at each time slot based on the state $s^t$ and action $a^t$. To guide the agent toward mission objectives, we designed a multi-component reward function that balances minimizing mission duration, ensuring collision avoidance, and promoting exploration. The reward $r^t$ comprises the following components.
\begin{enumerate}
	\item $r^t_1$: Collision penalty. For an action $a^t$, if the DT predicts that its execution would result in the UAV colliding with a building or exiting the mission area, the action is canceled, and a penalty is imposed, given by
	\begin{equation}
		r^t_1 = -w_1,
	\end{equation}
	where $w_1>0$ is a constant coefficient.
	\item $r^t_2$: Velocity penalty. To ensure the UAV reaches the target GU quickly, the UAV must maintain an appropriate velocity. Consequently, a penalty is applied when the velocity falls below a threshold, defined as
	\begin{equation}
		r^t_2 = -w_2,\ \text{if } v^t_u< J_{v,r} \cdot v_{max},
	\end{equation}
	where $w_2>0$ is a constant coefficient, and $J_{v,r} \in (0,1]$ represents the velocity threshold.
	\item $r^t_3$: Time penalty. To minimize mission duration,  the UAV incurs a penalty at each time slot, which increases with time, expressed as
	\begin{equation}
		r^t_3 = -w_3\cdot t,
	\end{equation}
	where $w_3>0$ is a constant coefficient.
	\item $r^t_4$: Service reward. To maximize data transmission rates thus shorten mission time, a reward is assigned for serving GUs, given by
	\begin{equation}
		r^t_4 = w_4 \cdot R^t_{u,k} + b_4,
	\end{equation}
	where $w_4>0$ and $b_4>0$ are constant weight and bias, respectively.
	\item $r^t_5$: Distance reward. The UAV selects the first unserved GU in the user scheduling vector as the target GU. A large reward is given when the distance to the target GU decreases significantly. To address obstacle avoidance, a smaller reward is given for slight distance reductions or minor increases, while a penalty is applied for substantial distance increases, defined as
	\begin{equation}
		r^t_5 =
		\begin{cases} 
			w_{5,1} \cdot |\Delta d^t_{u,tar}|, & \text{if } \Delta d^t_{u,tar} > J_{d,r}, \\
			w_{5,2} \cdot |\Delta d^t_{u,tar}|, & \text{if } -J_{d,r} \leq \Delta d^t_{u,tar} \leq J_{d,r}, \\
			- w_{5,1} \cdot |\Delta d^t_{u,tar}|, & \text{if } \Delta d^t_{u,tar} < -J_{d,r},
		\end{cases}
	\end{equation}
	where $w_{5,1}>w_{5,2}>0$ are constant coefficients, $J_{d,r}$ represents the distance reward threshold, and $\Delta d^t_{u,tar}$ denotes the distance change to the target GU at slot $t$.
	\item $r^t_6$: Exploration Reward. To discourage the UAV from continuously hovering in safe areas due to collision avoidance and to reduce prolonged flight time caused by repeatedly traversing already explored paths, we introduce an exploration reward mechanism, given by
	\begin{equation}
		r^t_6 =
		\begin{cases}
			w_6 \cdot \Delta S^t_u, & \text{if } \Delta S^t_u > 0, \\
			-w_6 , & \text{if } \Delta S^t_u = 0,
		\end{cases}
	\end{equation}
	where $w_6>0$ is a constant coefficient, and $\Delta S^t_u$ denotes the change in the number of grid cells newly covered by the UAV's communication range at slot $t$.
	\item $r^t_7$: Completion reward. Once the UAV complete the service for all users, a significant reward is assigned, defined as
	\begin{equation}
		r^t_7 = w_7 - t_f,
	\end{equation}
	where $w_7>0$ is a constant, and $t_f$ denotes the time slot at which the mission is completed.
\end{enumerate}

Furthermore, to balance the contributions of individual rewards, accelerate training convergence, and emphasize mission completion, the reward $r^t$ is defined as
\begin{equation}
	r^t=
	\begin{cases}
		g(r^t_{s}) + r^t_7, & \text{if } t = t_f, \\
		g(r^t_{s}) , & \text{if }  t \neq t_f,
	\end{cases}
\end{equation}
where $	g(r^t_{s})=\frac{2}{1 + \exp(-r^t_{s})}-1$ denotes the shaped reward function, and $r^t_{s}=r^t_1+r^t_2+r^t_3+r^t_4+r^t_5+r^t_6$.

\subsubsection{\textbf{State transition probability}}
The state transition probability represents the likelihood of transitioning to the next state $s^{t+1}$ when the agent executes action $a^t$ in state $s^t$. It is expressed as $P(s^{t+1} | s^t, a^t)$.

\subsection{TD3-based Trajectory Design}
Note that the UAV begins to navigate without prior environmental knowledge, with both the state space $\mathcal{S}$ and action space $\mathcal{A}$ being continuous and infinite. Therefore, a policy-based, model-free DRL algorithm is suitable for its trajectory design. Conventional candidates include policy gradient (PG), actor-critic, deterministic policy gradient, and DDPG \cite{10938344, 9708435, silver2014deterministic, lillicrap2015continuous}. Although DDPG integrates the strengths of PG and DQN by incorporating deep neural networks into the Q-learning framework, it often overestimates Q-values, which can impair effective policy learning \cite{fujimoto2018addressing}. In comparison, another advanced DRL algorithm, TD3, mitigates Q-value overestimation and enhancing performance through three key techniques.
%clipped double-Q learning, target policy smoothing, and delayed policy updates, as detailed below
\begin{enumerate}
	\item Clipped double-Q learning: TD3 employs two networks for both the online-critic and target-critic networks, generating two Q-values, respectively. By selecting the smaller Q-value as the target Q, the target-critic network is more accurately constructed, effectively reducing the risk of overestimation.
	\item Target policy smoothing: TD3 introduces noise to the actions produced by the target-actor network and smooths the corresponding Q-values, enhancing policy robustness against inaccuracies in Q-value estimates
	\item Delayed policy updates: TD3 updates the actor and target-actor networks less frequently than the critic and target-critic networks, allowing the critic to stabilize and reducing errors in policy updates.
\end{enumerate}

The TD3-based trajectory design framework, illustrated in the right side of Fig. \ref{fig:TD_TD3}, comprises an Actor network and a Critic network. The Actor network includes an online-actor network $\pi_{\phi}(\cdot)$ with parameters $\phi$, which approximates the agent's policy and generates actions, and a target-actor network $\pi_{\phi^\prime}(\cdot)$ with parameters $\phi^\prime$, which produces the target policy. The Critic network includes two online-critic networks, $Q_{\theta_1}(\cdot)$ and $Q_{\theta_2}(\cdot)$ with parameters $\theta_1$ and $\theta_2$, respectively, estimating action-value functions and selecting the smaller as the action-value, and two target-critic networks, $Q_{\theta_1^\prime}(\cdot)$ and $Q_{\theta_2^\prime}(\cdot)$, with parameters $\theta_1^\prime$ and $\theta_2^\prime$, generating the target Q-value by taking the smaller of the two computed Q-values.

During the training phase, the DT server generates a state based on the created VE and passes it to the online-actor network, which deterministically maps this state to an action. To balance exploration and exploitation, random noise is added to the action, expressed as
\begin{equation}
	a^t = \text{clip}\left(\pi_{\phi}(s^t)+\epsilon, a_{min}, a_{max} \right),
\end{equation}
where $\epsilon\sim\mathcal{N}\left(0,\sigma^2_n\right)$ represents zero-mean AWGN with variance $\sigma^2_n$. The clip function constrains actions within the bounds $[a_{min},a_{max}]$, where $a_{min}=\{0,0,0\}$ and $a_{max}=\{v_{max},\pi,2\pi\}$ define the lower and upper bounds of the actions, respectively. The agent then executes $a^t$, receives reward $r^t$, observes the next state $s^{t+1}$, and store the tuple $\langle s^t, a^t, r^t, s^{t+1}\rangle$ in the replay buffer $\mathcal{M}$.

Once sufficient tuples are collected, a mini-batch of size $M_b$ is uniformly sampled to update the networks. For each tuple $\langle s^t, a^t, r^t, s^{t+1}\rangle$, target policy smoothing mechanism is applied by computing a smoothed target action $a^{t+1}$ using the target-actor network, expressed as
\begin{equation}\small
	a^{t+1} = \text{clip}\left(\pi_{\phi^\prime}\left(s^{t+1}\right) + \text{clip}\left(\epsilon^\prime, -\sigma_{max}^2, \sigma_{max}^2\right), a_{min}, a_{max}\right),
\end{equation}
where $\epsilon^\prime \sim \mathcal N(0,\sigma_{n^\prime}^2)$ denotes zero-mean AWGN with variance $\sigma_{n^\prime}^2$, and $\sigma_{max}^2$ represents the maximum exploration noise.

The target Q-value is computed using the clipped double-Q learning technique. Specifically, the two target-critic networks evaluate the smoothed target action $a^{t+1}$ for the next state $s^{t+1}$, producing two Q-values $Q_{\theta_1^\prime}(s^{t+1},a^{t+1})$ and $Q_{\theta_2^\prime}(s^{t+1},a^{t+1})$, respectively. The smaller of these two values is used to calculate the target value, expressed as
\begin{equation}
	  y^t = r^t + \gamma \min_{i=1,2} Q_{\theta_i^\prime}(s^{t+1}, a^{t+1}),
\end{equation}
where $\gamma \in \left(0,1\right)$ denotes the discount factor that balances immediate and future rewards.

The online-critic networks, $Q_{\theta_1}(\cdot)$ and $Q_{\theta_2}(\cdot)$, are updated by minimizing the temporal-difference (TD) error between the predicted Q-values and the target value. The loss function for each online-critic network is defined as
\begin{equation}
	L(\theta_i) = \frac{1}{M_b} \sum_{t=1}^{M_b} \left( Q_{\theta_i}(s^t, a^t) - y^t \right)^2, \quad i=\{1,2\}.
\end{equation}
The parameters $\theta_1$ and $\theta_2$ are optimized using gradient descent to minimize their respective losses.

The online-actor network $\pi_{\phi}(\cdot)$ is updated to maximize the expected Q-value of its actions. The deterministic policy gradient is computed based on the Q-value from the first online-critic network, $Q_{\theta_1}$, denoted as
\begin{equation}
	\nabla_{\phi}J(\phi)=\frac{1}{M_b}\sum_{t=1}^{M_b}\nabla_{a}Q_{\theta_{1}}(s^t, a)|_{a=\pi_{\phi}(s^t)}\nabla_{\phi}\pi_{\phi}(s^t),
\end{equation}
where $J(\phi)$ represents the expected return under policy $\pi_{\phi}(\cdot)$. To implement delayed policy updates, the actor and target-actor networks are updated every $d_e$ iterations of the critic network updates.

The target networks, including the target-actor network $\pi_{\phi^\prime}(\cdot)$ and the target-critic networks $Q_{\theta_1^\prime}(\cdot)$ and $Q_{\theta_2^\prime}(\cdot)$, are updated using a soft update mechanism to ensure smooth convergence, expressed as
\begin{equation}
	\begin{cases}\theta_{i}^{^{\prime}}\leftarrow\tau\theta_{i}+(1-\tau)\theta_{i}^{^{\prime}}, \quad i=\{1,2\},\\\phi^{^{\prime}}\leftarrow\tau\phi+(1-\tau)\phi^{^{\prime}},&\end{cases}
\end{equation}
where $\tau \in (0,1)$ represents the update rate.

\begin{algorithm}
	\setstretch{0.92}
	\SetKwData{Left}{left}\SetKwData{This}{this}\SetKwData{Up}{up}
	\SetKwFunction{Union}{Union}\SetKwFunction{FindCompress}{FindCompress}
	\SetKwInOut{Input}{Input}\SetKwInOut{Output}{Output}\SetKwInOut{Initialize}{Initialize}
	
	\Input{UAV start position $\mathbf{w}^{0}_u$, positions of all GUs $\mathbf{w}_k$, $\forall k \in \mathcal{K}$, and user scheduling vector $\mathbf{g}_s$}
	\Output{Trained Actor $\pi_{\phi}(\cdot)$ and Critics $Q_{\theta_1}(\cdot)$, $Q_{\theta_2}(\cdot)$}
	\Initialize{Maximum number of episode $I^{TD3}_{max}$, maximum slots in each episode $T$, discount factor $\gamma$, policy delay $d_e$, mini-batch size $M_b$, online-actor network and online-critic network parameters $\phi$, $\theta_1$, $\theta_2$, the corresponding target network parameters $\phi^\prime \leftarrow \phi$, $\theta^\prime_1 \leftarrow \theta_1$, $\theta^\prime_2 \leftarrow \theta_2$, replay buffer $\mathcal{M}$}
	\For{$episode = 1$ \KwTo $I^{TD3}_{max}$}{
		Reset the environment, observe $s^1$\;
		\For{$t = 1$ \KwTo $T$}{
			$a^t = \text{clip}\left(\pi_{\phi}(s^t)+\epsilon, a_{min}, a_{max} \right)$\;
			Execute $a^t$, receive $r^t$, observe $r^t$, store $\langle s^t, a^t, r^t, s^{t+1}\rangle$ in $\mathcal{M}$\;
			\If{$|\mathcal{M}| \geq M_b$}{
				Sample mini-batch $\mathcal{M}_b \in \mathcal{M}$\;
				\ForEach{$\langle s^t, a^t, r^t, s^{t+1}\rangle$ $\in$ $\mathcal{M}_b$}{
					\small
					$\tilde\epsilon = \operatorname{clip}\!\bigl(\epsilon',-\sigma_{\max},\sigma_{\max}\bigr)$\;
					$a^{t+1} = \operatorname{clip}\!\bigl(\pi_{\phi'}(s^{t+1})+\tilde\epsilon,\;a_{\min},a_{\max}\bigr)$\;
					$y^t = r^t + \gamma\min_{i=1,2}Q_{\theta_i'}(s^{t+1},a^{t+1})$\;
				}
				Update online-critics $\theta_i \leftarrow \text{argmin}_{\theta_i}L(\theta_i), i=\{1,2\}$\;
			
			\If{$t \ \mathbf{mod}\ d_e = 0$}{
				Update online-actor via deterministic policy gradient on $J(\phi)$\; 
				Update target networks
				$\theta_{i}^{^{\prime}}\leftarrow\tau\theta_{i}+(1-\tau)\theta_{i}^{^{\prime}}, \quad i=\{1,2\}$,\
				$\phi^{^{\prime}}\leftarrow\tau\phi+(1-\tau)\phi^{^{\prime}}$\;
				}
			}
		}	
	}
	\caption{TD3-based trajectory design}
	\label{alg:TD3_algorithm}
\end{algorithm}

Algorithm \ref{alg:TD3_algorithm} outlines the TD3-based trajectory design process for the UAV. It takes as input the start position of the UAV, positions of all GUs and the user scheduling vector, and outputs well-trained actor and critic networks with optimal parameters. Initially, hyperparameters, network parameters, and a replay buffer $\mathcal{M}$ are initialized. In each episode, the environment is reset to observe the initial state $s^1$ (line 2). For each time slot, the UAV selects an action $a^t$ by adding noise to the online-actor's output, executes it, observes the reward $r^t$ and the next state $s^{t+1}$, and store the tuple $\langle s^t, a^t, r^t, s^{t+1}\rangle$ in $\mathcal{M}$ (lines 4-5). Once sufficient tuples are collected, a mini-batch of size $M_b$ is sampled to compute smoothed target actions and target Q-values using clipped double-Q learning. The online-critic networks are updated by minimizing the TD error, while the online-actor and target networks are updated periodically via policy gradient and soft updates, respectively, ensuring stable and robust policy learning (lines 6-15).

\section{Simulation and Discussion}\label{Sec.Simulation}

\subsection{Simulation Settings}
The simulation scenario for the UAV-based LAWN deployment system is set as a 3D space with dimensions of $1000\ \text{m} \times 1000\ \text{m} \times 150\ \text{m}$. Within the mission area, previously unknown buildings are randomly distributed and act as signal-blocking obstacles. These buildings are modeled as square or rectangular structures, with lengths and widths drawn from random distributions, and heights following a Rayleigh distribution with a scale parameter of 100, bounded between $40\ \text{m}$ and $150\ \text{m}$ \cite{series2013propagation}. GUs are randomly distributed on the ground outside the buildings. Each GU requires a data volume of $\Lambda_k = 10\ \text{Mbits}$. The UAV is constrained to fly at altitudes of up to $140\ \text{m}$, and can travel at a maximum speed of $40 \ \text{m/s}$. The duration of each sub-slot is set as $\delta_1 = 0.5 \ \text{s}$, $\delta_2 = 0.5 \ \text{s}$, and $\delta_3 = 0.2 \ \text{s}$, respectively. For data transmission, the BS and UAV transmit data with powers of $p_b=40\ \text{dBm}$ and $p_u = 10\ \text{dBm}$, respectively, over a $10\ \text{MHz}$ bandwidth. The carrier frequency is set as $f_c = 2\ \text{GHz}$, and the noise power is set as $N_0 = -75\ \text{dBm}$. The UAV's sensing and communication angles are set as $\beta_{sen}=\pi/6$ and $\beta_{com} = \pi/4$, respectively. For channel propagation, large-scale fading is modeled with propagation loss factors of $\gamma_{\mathrm{LoS}} = 0.1\ \text{dB}$ for LoS links and $\gamma_{\mathrm{NLoS}} = 21\ \text{dB}$ for NLoS links, as specified in \cite{al2014optimal}. For small-scale fading, LoS links are modeled using Rician fading with a Rician factor of 15 dB, while NLoS links are modeled using a Rayleigh distribution \cite{wang2021trajectory}.

The user scheduling algorithm is configured with an initial temperature $T_0=2000$, a cooling rate $C_r=0.998$, and a maximum iteration count $I^{SA}_{\max}=4000$. For the neighborhood operators $\mathcal{O}=\{\text{swap}, \text{insert-move}, \text{reverse-subsequence}\}$, the corresponding bias coefficients are $\beta_o=\{3,2,1\}$ and the sensitivity coefficients are $\alpha_o=\{1.0, 1.1, 1.2\}$. In the MDP formulation, the state dimensions are defined as $m=100$ and $n=100$. Based on the user scheduling vector, the attractive constant $\mu_k$ for the highest-priority GU starts at $K$, decreasing progressively for each lower-priority GU. The GU potential distance threshold is $J_{d,p}=50\ \text{m}$. For the reward function, the collision penalty coefficient is $w_1=100$ to prioritize collision avoidance. The velocity penalty uses a threshold $J_{v,r}=0.6$ to allow speed reduction near obstacles, with $w_2=3$. A small time penalty coefficient $w_3=0.01$ balances mission duration without dominating other objectives. Service reward parameters $w_4=2$ and $b_4=1$ encourage efficient data transmission to GUs. Distance reward coefficients $w_{5,1}=3$ and $w_{5,2}=1$, with a threshold $J_{d,r}=20$, promote progress toward the target GU. The exploration reward coefficient is $w_6=0.5$ to incentivize new area coverage. Finally, a completion reward coefficient $w_7=500$ emphasizes mission completion.

\begin{table}[htbp]
	\centering
	\caption{\normalsize Actor Network Architecture}
	\small % 或用 \footnotesize 缩小字号
	\begin{tabular}{ll}
		\toprule
		\textbf{Module} & \textbf{Core Components} \\
		\midrule
		\textbf{CNN 1 ($\mathbf{S}^t_1$)} & Conv2d ×2: 1$\rightarrow$128$\rightarrow$256 \\
		\textbf{CNN 2 ($\mathbf{S}^t_2$)} & Conv2d ×2: 1$\rightarrow$128$\rightarrow$256 \\
		\textbf{Feature Fusion} & Concat → Conv2d ×2: 512$\rightarrow$1024$\rightarrow$2048 \\
		\textbf{Attention Module} & Channel Attention + Spatial Attention \\
		\textbf{Global Pooling} & AdaptiveAvgPool2d → (batch, 2048) \\
		\textbf{Shared FC Layers} & FC: 2048$\rightarrow$4096$\rightarrow$2048 \\
		\textbf{Velocity} & FC: 2048$\rightarrow$512$\rightarrow$128$\rightarrow$1 \\
		\textbf{Vertical Angle} & FC: 2048$\rightarrow$512$\rightarrow$128$\rightarrow$1 \\
		\textbf{Horizontal Angle} & FC: 2048$\rightarrow$512$\rightarrow$128$\rightarrow$1 \\
		\bottomrule
	\end{tabular}
	\label{tab:actor_arch_2col}
\end{table}

Table \ref{tab:actor_arch_2col} summarizes the key components of the Actor network. It features dual convolutional branches for extracting GU and obstacle features, followed by feature fusion and attention modules. After global average pooling and shared fully connected (FC) layers, the network is divided into three branches to generate velocity, vertical angle, and horizontal angle, respectively. The Critic network has a similar architecture to the Actor network; however, it concatenates the global pooled features with the action input before passing them through shared FC layers and a single branch to output the Q-value. For training, the model uses the Adam optimizer with a learning rate of $0.0001$. A replay buffer of size $300000$ stores past experiences, and mini-batches of $256$ are sampled for updates. The discount factor is set as $0.99$, and each episode lasts up to $500$ slots. The target network is updated every $2$ slots to enhance training stability.

\subsection{Performance Evaluation and Comparison}
\begin{figure}[htbp]
	\centering
	\includegraphics[width=0.45\textwidth]{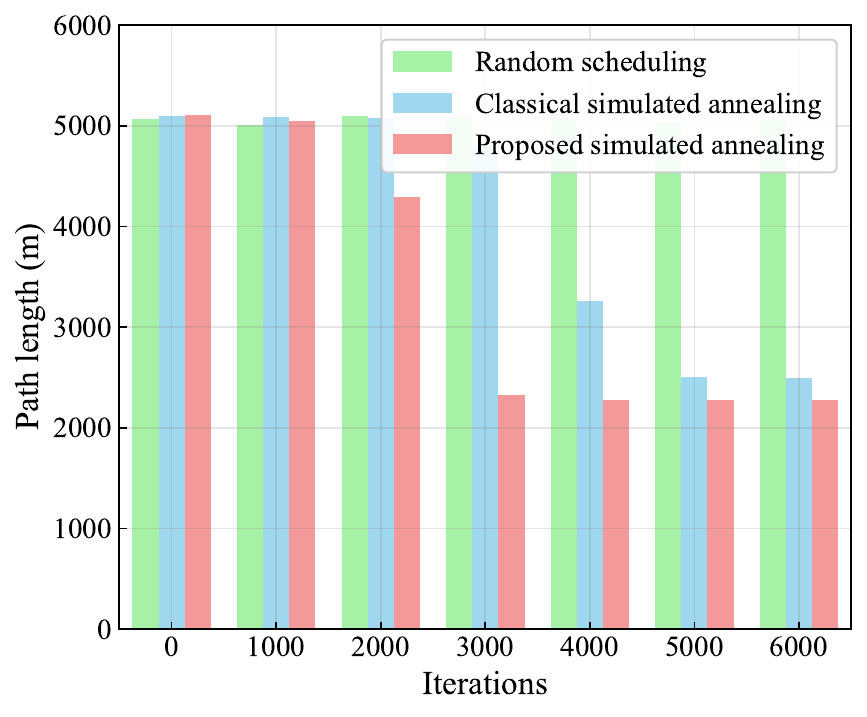}
	\caption{Comparison of the path length over iterations for random scheduling, classical annealing, and the proposed algorithm.}
	\label{fig. SA_comparison}
	\vspace{0 cm}
\end{figure}
In the context of user scheduling, Fig. \ref{fig. SA_comparison} illustrates the convergence performance of random scheduling, the classic simulated annealing algorithm, and the proposed scheduling approach in deriving the scheduling order of 10 GUs without considering obstacles. The classical method, initialized with a random schedule and relying solely on swap operations, improves slowly and yields suboptimal results. By contrast, our approach, with nearest-greedy initialization and adaptive neighborhood exploration, converges faster and achieves the shortest path, demonstrating its superior efficiency and solution quality.

\begin{figure}[htbp]
	\centering
	\hspace{-0.5cm}
	\subfloat[2D trajectory]{
		\includegraphics[width=0.21\textwidth]{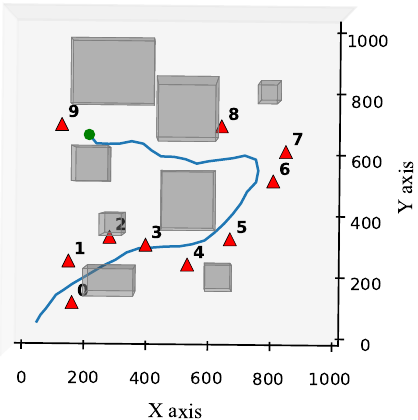}}
	\hspace{0\textwidth} 
	\subfloat[3D trajectory]{
		\includegraphics[width=0.27\textwidth]{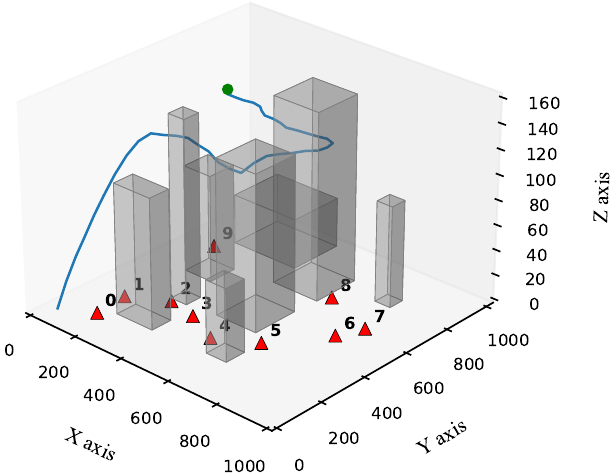}}
	\caption{The UAV's maneuver trajectory on 2-D and 3-D views with the proposed DT-assisted SATD3TD.}
	\label{UAVpath}
\end{figure}

To directly illustrate the effectiveness of the proposed algorithm, Fig. \ref{UAVpath} depicts the UAV trajectory in a scenario involving 10 GUs and 8 buildings, where the UAV's initial point is at the lower-left corner. The red triangles represent the served GUs, with the numbers above them indicating the scheduling order obtained from the algorithm in Section \ref{user scheduling}. Smaller numbers correspond to higher service priority. The blue curve denotes the UAV’s flight path, while the green circle marks the mission completion point. By analyzing the trajectory, it can be observed that the UAV is able to avoid obstacles and quickly ascend to a safe altitude. Furthermore, it efficiently completes service to all GUs in the scheduled order without colliding with any buildings. Even in an unknown environment, the UAV maintains an orderly flight path, avoiding unnecessary maneuvers while ensuring rapid service delivery to all GUs, thereby validating the effectiveness of the proposed method.

In order to evaluate the performance of DT-assisted SATD3TD, several baseline algorithms are employed for comparison, including (1) the SATD3TD algorithm without incorporating the DT technique \cite{wang2021trajectory}, denoted as ``SATD3TD without DT"; (2) a trajectory design algorithm that employs DDPG after the scheduling phase in Section \ref{user scheduling} within the DTTDF, denoted as ``DT-assisted SADDPG"; and (3) the SADDPG algorithm without the DT technique \cite{9652887}, denoted as ``SADDPG without DT". Specifically, the ``without DT" refers to a setting in which the UAV cannot perform trajectory design based on the reconstructed environment. Instead, it relies solely on real-time sensing data available at each slot for decision-making.

\begin{figure}[htbp]
	\centering
	\includegraphics[width=0.45\textwidth]{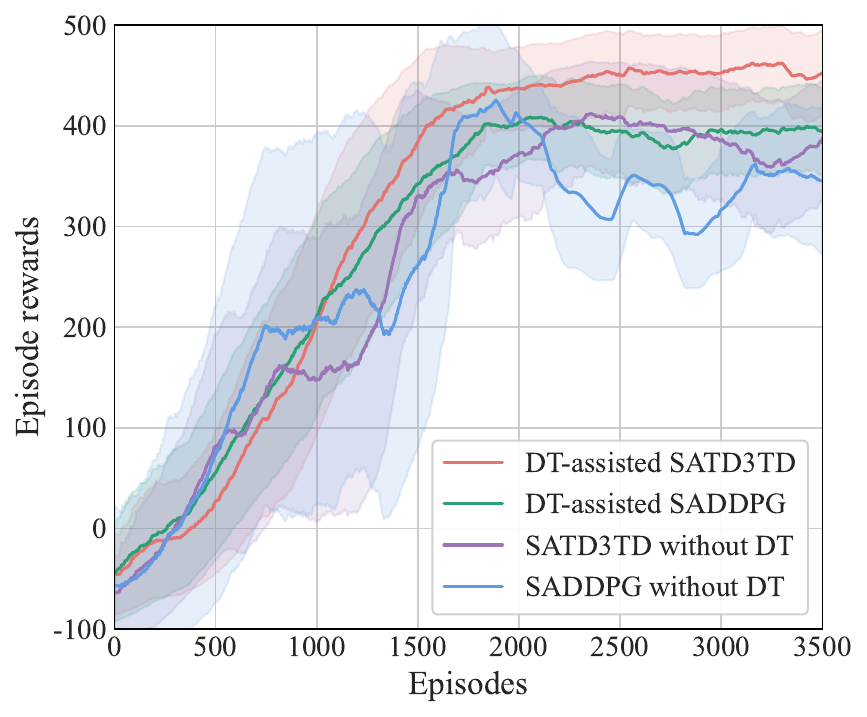}
	\caption{Comparison of reward convergence performance of SATD3TD and SADDPG training with and without DT. The shaded regions represent half the standard deviation computed using a sliding window of size 20. All curves are uniformly smoothed for visual clarity.}
	\label{fig. reward_vs_eposide}
	\vspace{0 cm}
\end{figure}

\begin{figure}[htbp]
	\centering
	\hspace{-0.5cm}
	\subfloat[Number of served GUs of episodes]{
		\includegraphics[width=0.24\textwidth]{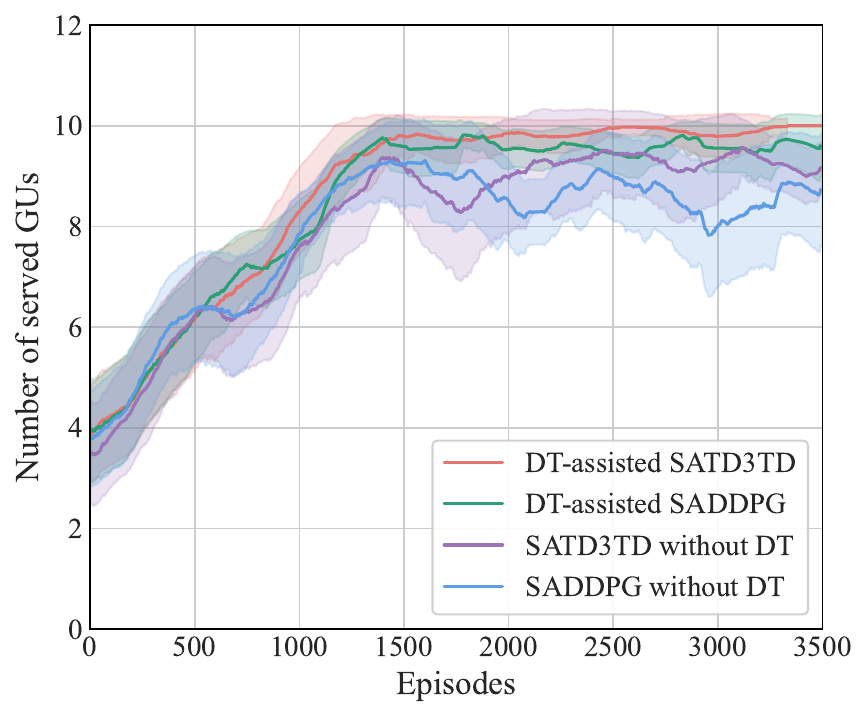}}
	\hspace{0\textwidth} 
	\subfloat[Number of collisions of episodes]{
		\includegraphics[width=0.24\textwidth]{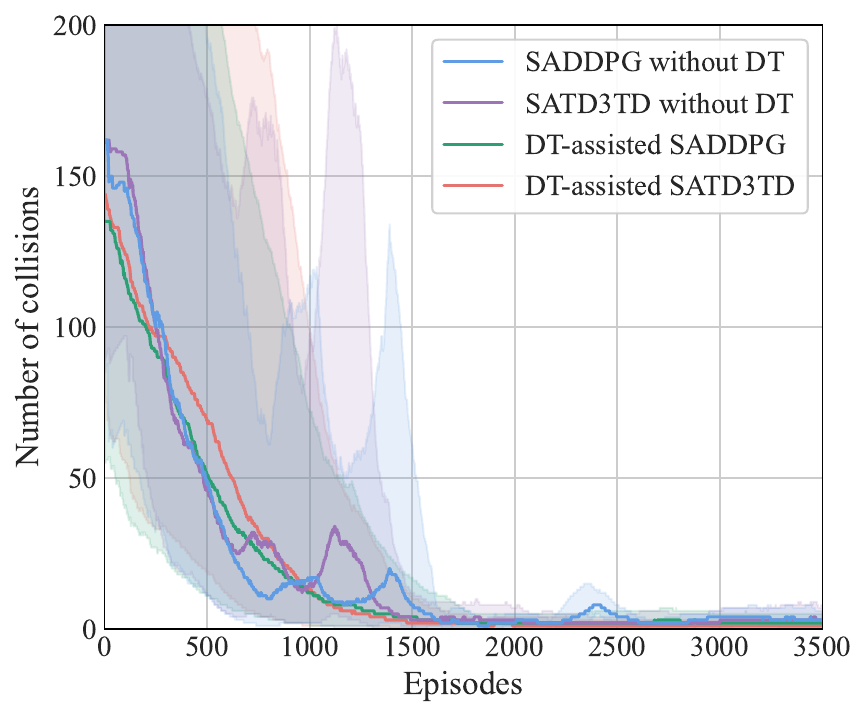}}
	\caption{Comparison of convergence performance in terms of the number of served GUs and collisions for SATD3TD and SADDPG training with and without DT. The shaded regions represent half the standard deviation computed using a sliding window of size 20. All curves are uniformly smoothed for visual clarity.}
	\label{fig. serve_collide_eposide}
\end{figure}

The convergence performance of DT-assisted SATD3TD and various baseline methods is illustrated in Fig. \ref{fig. reward_vs_eposide} and Fig. \ref{fig. serve_collide_eposide}. As depicted in Fig. \ref{fig. reward_vs_eposide}, although all methods eventually converge, those incorporating DT achieve faster convergence to higher episode rewards and exhibit superior stability, as evidenced by smoother curves and narrower variance bands. In contrast, non-DT methods demonstrate greater volatility and slower learning progress, reflected in larger fluctuations throughout training. These improvements arise from DT’s capability to reduce environmental uncertainty through a continuously updated VE, which provides more reliable state representations and feedback for model training. As a result, the UAV can focus its learning on promising trajectories rather than wasting effort on uninformative exploration. The effectiveness of DT is further validated in Fig. \ref{fig. serve_collide_eposide}. Specifically, Fig. \ref{fig. serve_collide_eposide}(a) shows that DT-enhanced methods reach the maximum number of served GUs more rapidly and reliably, benefiting from improved trajectory adaptation based on real-time environmental feedback and dynamic GU updates in the DT. Meanwhile, Fig. \ref{fig. serve_collide_eposide}(b) highlights DT’s crucial role in ensuring safety, as its real-time building reconstruction enables the UAV to adjust its path proactively, leading collision rates to drop quickly to near zero and remain consistently low.

\begin{figure}[htbp]
	\centering
	\includegraphics[width=0.45\textwidth]{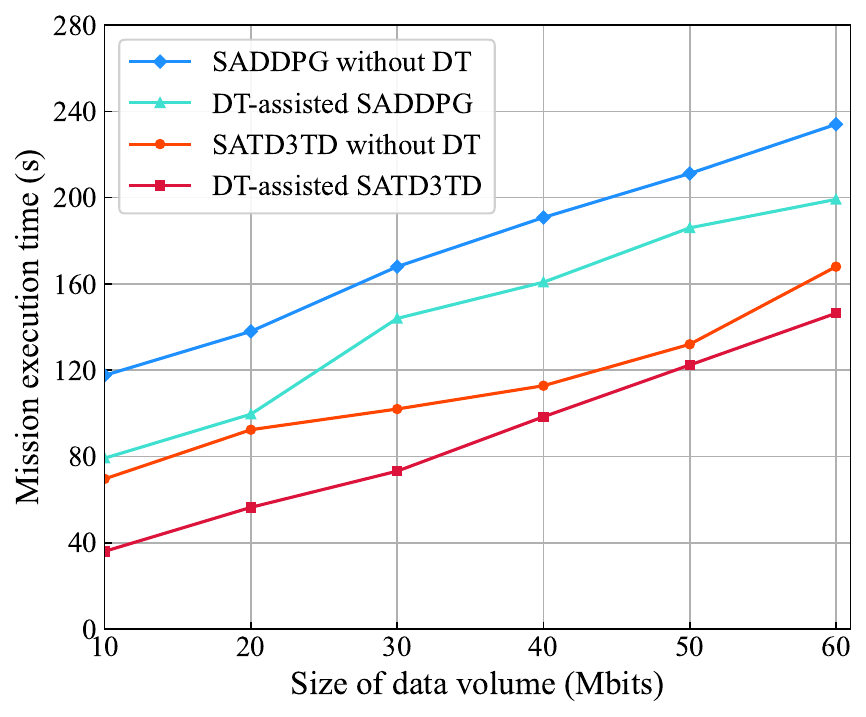}
	\caption{Comparison of the mission execution time under different data size volumes for SATD3TD and SADDPG with/without DT.}
	\label{fig. datasize_vs_missiontime}
	\vspace{0 cm}
\end{figure}

In fig. \ref{fig. datasize_vs_missiontime}, we evaluate the mission execution times under varying data volumes for the SATD3TD and SADDPG with and without DT integration. All curves exhibit an upward trend, indicating that increased data volume leads to longer mission durations. Notably, the integration of DT techniques consistently enhances performance for both SATD3TD and SADDPG, resulting in significantly shorter execution times compared to their non-DT counterparts. These efficiency gains arise from the precise VE provided by the DT, which enables the UAV to accurately navigate toward LoS links and avoid A2G blockages. Moreover, TD3-based methods consistently outperform their DDPG-based counterparts, owing to their more reliable policy updates and reduced value estimation bias. Among all methods, DT-assisted SATD3TD achieves the shortest mission execution times compared to other algorithms, demonstrating its strong capability to handle increasing data volume efficiently.

\begin{figure}[htbp]
	\centering
	\includegraphics[width=0.45\textwidth]{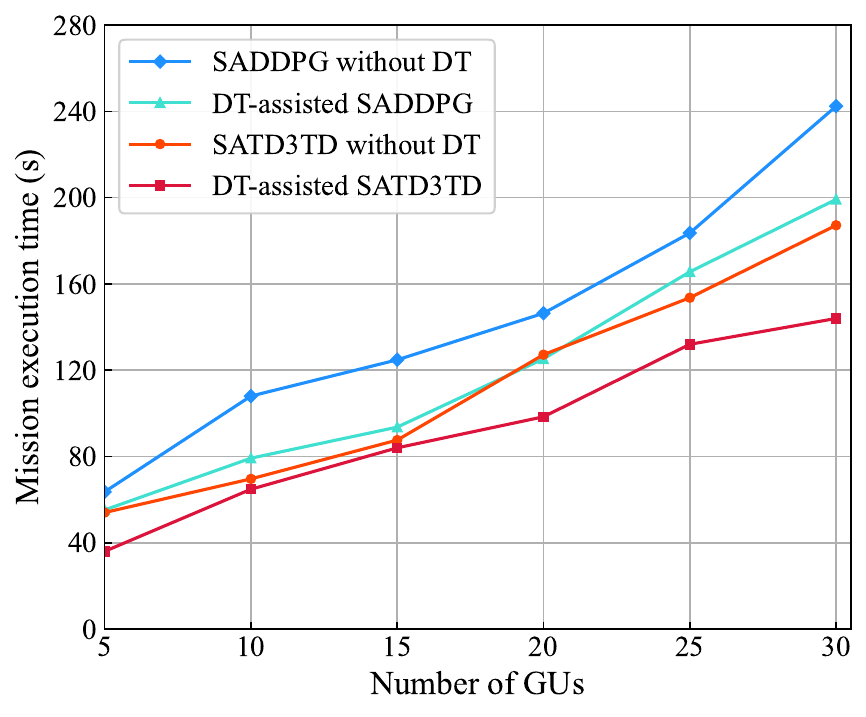}
	\caption{Comparison of the mission execution time under different number of GUs for SATD3TD and SADDPG with/without DT.}
	\label{fig. GUnumber_vs_missiontime}
	\vspace{0 cm}
\end{figure}

Fig. \ref{fig. GUnumber_vs_missiontime} depicts the mission execution times under varying numbers of GUs. As expected, all curves rise with increasing GU numbers, highlighting the fundamental challenge of scaling communication services to accommodate more users. The presence of DT further improves the effectiveness and stability of both SATD3TD and SADDPG. Specifically, DT-assisted SATD3TD achieves the shortest mission times, while DT-assisted SADDPG reaches performance comparable to SATD3TD without DT, despite DDPG’s inherent instability and learning limitations. This improvement is attributed to DT’s ability to generate rich and accurate state information, effectively compensating for these shortcomings. Moreover, DT-assisted SATD3TD demonstrates superior scalability, as evidenced by its relatively flatter mission time growth curve with increasing GUs, reflecting its robust capability to efficiently manage a larger user base through DT-assisted dynamic decision-making.

\begin{figure}[htbp]
	\centering
	\includegraphics[width=0.45\textwidth]{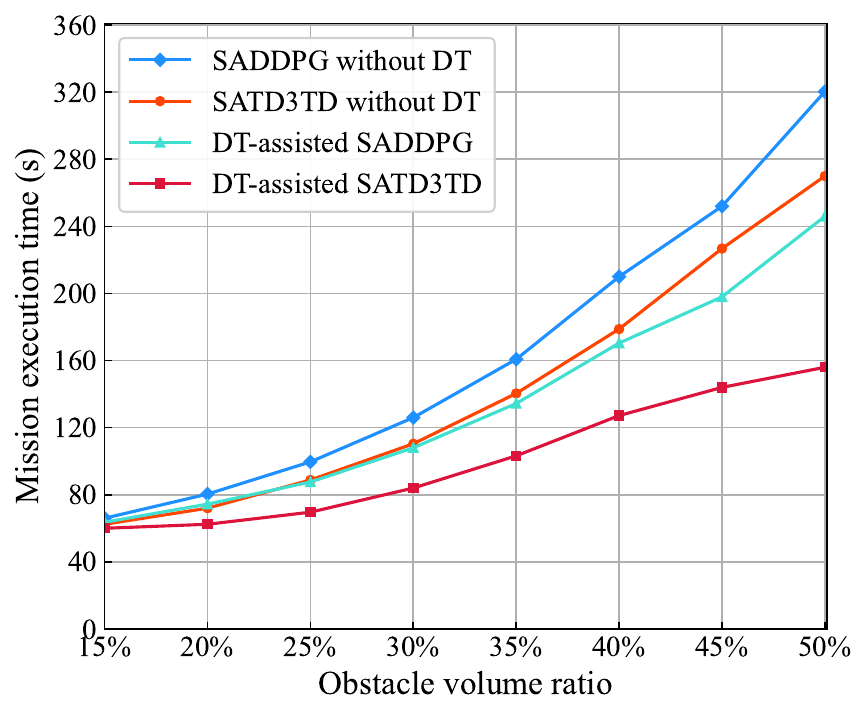}
	\caption{Comparison of the mission execution time under different obstacle volume ratios for SATD3TD and SADDPG with/without DT.}
	\label{fig. obstacle_vs_missiontime}
	\vspace{0 cm}
\end{figure}

Fig. \ref{fig. obstacle_vs_missiontime} shows the impact of obstacle volume ratio on mission execution time, where the ratio is defined as the total volume of buildings divided by the volume of the 3D mission space. As the ratio increases, all methods require more time, reflecting the additional complexity of path planning and obstacle avoidance in denser environments. Among them, DT-assisted SATD3TD consistently achieves the shortest execution time across the entire range. While DT-assisted SADDPG and SATD3TD without DT exhibit similar performance at lower obstacle ratios, DT-assisted SADDPG gradually outperforms SATD3TD as the environment becomes more complex.  Notably, DT-based algorithms show a smaller increase in mission time compared to non-DT counterparts, highlighting DT's ability to mitigate the challenges of higher building density. This advantage stems from DT's continuous environmental reconstruction, which provides accurate, real-time data, enabling DT-based algorithms to adapt more effectively and maintain superior efficiency in complex settings.

\section{Conclusion}\label{Sec.Conclusion}
In this paper, we have proposed trajectory design for UAV-based LAWN deployment in completely unknown environments, where the UAV transmitted ISAC signals to deliver communication services to GUs while collecting echoes for VE reconstruction on the DT server. These reconstructions guided the UAV’s flight decisions. The application of DTTDF can significantly accelerate model training, improved deployment performance, and ensured UAV safety during mission execution. By formulating the trajectory design problem as an MDP and integrating the simulated annealing algorithm with the TD3 framework, we have developed an efficient and robust DT-assisted SATD3TD approach, which minimized mission execution time while guaranteeing collision avoidance. Simulation results have verified the effectiveness of the proposed method, demonstrating its superiority over existing algorithms.

\normalem
\bibliographystyle{IEEEtran}
\bibliography{transreference}
\begin{comment}
\newpage

\section{Biography Section}
If you have an EPS/PDF photo (graphicx package needed), extra braces are
 needed around the contents of the optional argument to biography to prevent
 the LaTeX parser from getting confused when it sees the complicated
 $\backslash${\tt{includegraphics}} command within an optional argument. (You can create
 your own custom macro containing the $\backslash${\tt{includegraphics}} command to make things
 simpler here.)
 
\vspace{11pt}

\bf{If you include a photo:}\vspace{-33pt}
\begin{IEEEbiography}[{\includegraphics[width=1in,height=1.25in,clip,keepaspectratio]{fig1}}]{Michael Shell}
Use $\backslash${\tt{begin\{IEEEbiography\}}} and then for the 1st argument use $\backslash${\tt{includegraphics}} to declare and link the author photo.
Use the author name as the 3rd argument followed by the biography text.
\end{IEEEbiography}
\end{comment}
\vspace{11pt}

\vfill

\end{document}